# Security risk assessment in Internet of Things systems

Jason R.C. Nurse [1], Sadie Creese [1], David De Roure [2]
[1] *Department of Computer Science, University of Oxford, UK*
[2] *Oxford e-Research Centre, University of Oxford, UK*

## Abstract

Cybersecurity risk assessment approaches have served us well over the last decade. They have provided a platform through which organisations and governments could better protect themselves against pertinent risks. As the complexity, pervasiveness and automation of technology systems increases however, particularly with the Internet of Things (IoT), there is a strong argument for the need for new approaches to assess risk and build trust. The challenge with simply extending existing assessment methodologies to these systems is that we could be blind to new risks arising in such ecosystems. These risks could be related to the high degrees of connectivity present, or the coupling of digital, cyber-physical and social systems. This article makes the case for new methodologies to assess risk in this context which consider the dynamics and uniqueness of IoT, but also the rigour of best practice in risk assessment.

## 1. Introduction

As technology continues to permeate modern-day society, the security of, and trust that we place in, these systems becomes an increasingly significant concern. This is particularly given the plethora of attacks being launched that target organisations, governments and society. The traditional approach to address such challenges has been to conduct cybersecurity risk assessments that seek to identify critical assets, the threats they face, the likelihood of a successful attack, and the harms that may be caused. Only in this way, and after the identified risks have been prioritised, would appropriate approaches be selected to effectively address them.

The Internet of Things (IoT) is set to benefit society through a range of smart platforms and a pervasive coupling of digital, cyber-physical and social systems. This coupling allows relationships between systems that may vary drastically in terms of density, time, and automation. The challenge with IoT from a security and trust management perspective however, is that existing risk assessment methodologies were established prior to it. And, as such, may not cater to the complexity or pervasiveness of these automated systems. Ultimately, adopting these methods to IoT may make us blind to new risks arising in their ecosystems. These may relate to cyber-attack, but equally to new social processes which emerge at the scale of the population in real-time (e.g., viral effects in social media), and to the "natural disasters" inherent in the accidental failure of IT systems.

In this article therefore, we carefully analyse the reasons why current risk assessment approaches are unsuitable for IoT, and highlight the need for new approaches to underpin trust in IoT-based systems. It is only by crafting such methods, in partnership with industry, government and academia, that we will be prepared to address the threats facing IoT.

## 2. The current cybersecurity risk assessment paradigm

### 2.1 Core concepts of risk assessment

Risk assessment is generally understood as the process of identifying, estimating and prioritising risks to the organisational assets and operations [1]. This is a critical activity within risk management as it provides the foundation for the identified risks to be treated. Treatment options include: risk acceptance for cases where the risk is at an acceptable level considering the organisation's risk appetite; risk mitigation using security controls; risk transfer through the purchase of cyber insurance; or risk avoidance by removing the affected asset. There are several core concepts that feature within risk assessment, such as assets, vulnerabilities, threats, attack likelihood, and impact or cyber-harm.

Assets can be defined as any items of value to the organisation, and can have various different properties. For instance, assets can be tangible (e.g., technical infrastructure) or intangible (reputation or a business process), or they can be small components within a system or be the system themselves. Vulnerabilities are the ways in which assets can be exploited, and define weaknesses in assets or in the risk controls put in place to protect them. A threat is the action that could adversely impact an asset, and typically involves exploiting a vulnerability. Such actions may be deliberate (e.g., stealing corporate data) or accidental (e.g., being the victim of a social engineering attack). Cyber risk is the combination of these concepts, and considers the likelihood of a successful threat or attack occurring, and the harms that may result to assets.

### 2.2 Approaches to risk assessment

Although the fundamental process behind cybersecurity risk assessment has been clearly defined, there is a reasonable degree of flexibility in how its sub-processes are implemented. This flexibility has resulted in the rise of several different methods, guides and tools for conducting risk assessments. These vary according to contexts as well as the type of organisations for which the assessment is designed. A few examples of the most popular and well-regarded of these approaches include NIST SP800-30, ISO/IEC 27001, OCTAVE, CRAMM and EBIOS [2], and their origins range from standard-setting bodies (e.g., NIST and ISO/IEC) to governments (e.g., CRAMM from the UK and EBIOS from France). These approaches are all periodically applied within organisations to assess risk.

Given the wide variety of risk assessment methodologies, instead of focusing on each one individually, a better approach to analysis is to consider the aspects that set them apart. Two of the most significant aspects of these are the nature of the approach and how it measures risk; these can be seen in recent survey work [3]. In terms of the nature of the approach, we specifically consider the fact that some risk assessment processes are grounded around critical assets and the harm that may occur to them, and others around the threats and how feasible they are. The NIST approach is one of the latter, and therefore, its first steps are to identify threat sources and events [1]. After this, it advocates identifying the vulnerabilities that might be exploited and the respective likelihood and impact of threat events, before then determining risks.

Other approaches such as OCTAVE, however, emphasise the identification of critical assets first, and then build outwards in terms of how those assets can be threatened, and the result of the threat [2]. From this process, an understanding of the risk is developed. The benefit of the asset-oriented approach is that it ensures assessments are centred on critical assets rather than ephemeral threats, while the threat-oriented approach tends to be better catered to current threat landscapes.

The way that risks are measured is also a heavily contested factor. With regards to the rating of a threat's likelihood and impact, qualitative measures – for instance, variations on high, medium and low – can be found in most of the popular approaches (e.g., NIST SP800-30, ISO/IEC 27001, OCTAVE). The benefit is the simplicity offered, both in setting risk appetites, measuring risks (through the combination of threat likelihood and impact ratings), and communicating risk information to others. The disadvantage with the qualitative approach however, is its subjectivity and lack of precision [3]. For instance, one person's view of a threat as low may not conform to another person's belief.

As a result, a host of techniques have been proposed to address such problems, with probabilistic models featuring in many of them. Although these manage to address some of the issues, they often raise other significant questions. The most common of which pertains to the complexity of the analysis (therefore, increased likelihood of being error-prone and difficult to communicate to others), and challenge in accurately estimating the probability of the threat event occurrence and value of the impact (given lack of sufficient data). These aspects have limited the application of quantitative analysis techniques generally, and there are few known cases of their utility or success in complex and highly interconnected systems. A similar point applies to the lack of rigorous dynamic risk assessment approaches – hence the prevalence of periodic assessment techniques.

Beyond the distinguishing factors mentioned above, there are a number of additional areas that help to characterise and inform risk assessment approaches, and are useful or our IoT context. Survey work [3] has highlighted: the extent to which the methodology accommodates for risk propagation or dependencies; how the various resources in the organisational infrastructure are valued and from what perspectives; and whether the approach prioritises reducing known system risks, or expanding analyses to future scenarios and postulating based on past experiences. Each of these has its own nuances and application scenarios.

## 3. The relevant dynamics of IoT

By its very nature, IoT is a complex technology paradigm. This complexity is portrayed in part in Figure 1, and through its various applications, from logistics and manufacturing, to healthcare and smart infrastructures. From a risk assessment and trust perspective, the dynamics of IoT is of particular interest for several reasons. In what follows, we complement our reflection on risk assessment by examining the dynamics of IoT – this sets the foundation for our core argument in Sections 4 and 5.

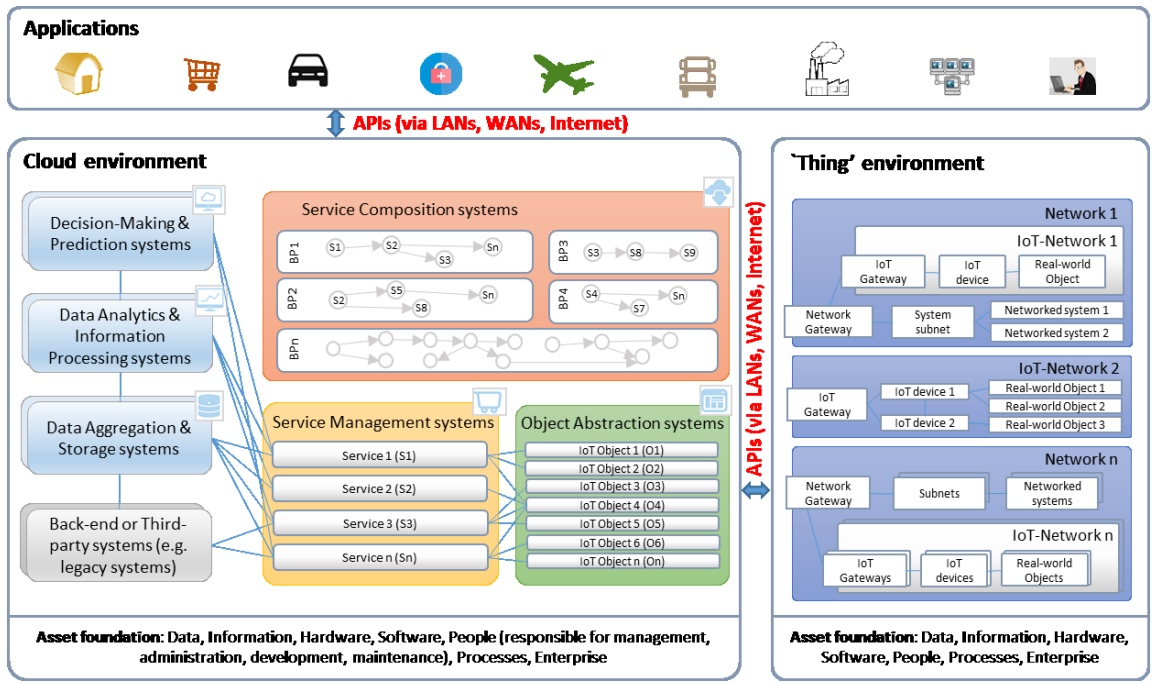

Figure 1: The array of components commonly featured in IoT systems and how they may be connected across Application, Cloud and 'Thing' environments (inspired, in part, by [4])

The first point of note regarding IoT is the variability of scale in devices and systems. One of the central advantages of IoT is its ability to expand (or shrink) in scale, and accommodate a wide range of new systems and "things" as shown in Figure 1. Indeed, the essence of the evolution of IoT is in the instrumentation of our environments in the broadest sense. We are witnessing the inclusion of digital functionality which typically enables remote control and collaboration, inside any and every aspect of the natural and constructed world – in a sense meaning that nothing is necessarily out-of-scope for future IoT.

Another aspect of IoT is its dynamism and the temporality of connections between devices [4]. IoT devices may be loosely coupled to perform some task and break connections once it is complete, or, connections may be persistent. It is important to understand the level of temporality required for a specific IoT context given the resulting impact on risk (e.g., persistence in connection from unauthorised devices). One final driving factor in the nature of the relationships will be the resources required to support the management and control activities for such relationships. Limited resources will mean that IoT devices may be forced to adopt regimes that allow for a small variety of relationships due to the resource required to maintain them; or, they could be coupled with cloud systems (Figure 1), which also would need to be assessed for risk.

The heterogeneity of actors capable of interacting within IoT ecosystems is also a significant characteristic. IoT devices are often accessible across organisations and may be uniquely addressable online. In instances where they allow loose coupling, there could be any number or type of actors, be they devices, people or systems, interacting with them (or, part of their asset foundation – Figure 1). While this is ideal from the perspective of IoT generally in allowing adaption to suit tasks, as we will discuss in Section 4, it has multiple disadvantages from a trust standpoint. This also raises issues for trust management given that the heterogeneity of actors and features of devices may mean that both benign and malicious relationships may form [5]. Moreover, as some relationships may be spontaneous or temporal, it can be challenging to track misbehaving actors, and also difficult to pinpoint the location or propagation of risks given that they may be distributed across various devices.

A factor that is often overlooked in discussions of IoT is the *glue* that binds these systems, especially those that are cyber-physical or cyber-social. If we reflect on the state of research and practice in the security and trust in IoT systems, we can see that there is a notable amount of work focused on device components and interfaces [5]. The reality is, however, that the process through which these devices are bound, and the connections that allow them to couple and operate, is also extremely important. This importance is driven by the central nature of these processes and connections, and the fact that there is arguably little emphasis on security and trust here. Some articles have even called for additional research efforts to tackle issues related to integration of IoT (though, focusing more on secure middleware) [5]. There remains little work considering the nature of the glue and how it binds across actors of such variable types.

## 4. Where current risk assessment methods fail within IoT

The dynamics of IoT systems will make risk assessment using current practices challenging. We deal here with some of the key tensions that must be overcome if we are to enable trust in IoT environments.

### 4.1 Shortcomings of periodic assessment

The periodic assessment of system risk is typically triggered by concerns that an organisation's prior assessment of risk may no longer be valid. Such triggers include significant change in the system, change in business processes, or threat intelligence providing insight into newly expected attacks. Of course, it is entirely possible, as we note above, that the assessment misses a risk that is later realised. Or, equally, that the re-assessment is not triggered and therefore risk is carried, and later materialises. This is not particular to IoT.

The reason why risk assessment approaches are inadequate for IoT is that their periodic assessment nature, which is already a notable weakness [6], is exacerbated due to the IoT dynamics outlined above. For instance, the variability in scale of IoT systems means that the probability of a new system emerging between periodic assessments will be very high. To be effective, risk assessment would need to be able to predict and consider the possible systems that might emerge prior to the next periodic assessment – this is extremely challenging, and current approaches typically do not mandate it [6]. Therefore, we might argue that a required extension to current practice for IoT would be to have an element of assessment for *potential* given the dynamics of devices currently in use, and those that could become connected.

### 4.2 Changing systems boundaries yet limited system knowledge

Risk assessment is currently focused on determining risk for systems that exist, and even now, there are challenges in developing a comprehensive understanding of such systems' environments [6]. We have noted that this is unlikely to be sufficient for IoT because systems may well change shape quicker that such assessments can account for periodically. Even if we are able to enhance current techniques to consider potential changes, we will still face the challenge of shifting system boundaries. The pace of change is potentially so high that we will be forced to manage risks with limited system knowledge. This is not outside the understanding of risk management in general, but it is outside the current practice in digital systems risk assessment.

For IoT, the risk community will have to develop ways of abstracting from systems' detail, and yet still properly assess the risk faced. It may be that this forces a harm-centric approach to risk identification in every assessment made. The difficulty that the professional community may face is that this will initially lead to the identification of many more potential risks, and the accusation of scare tactics since many will never materialise. This will undoubtedly force the adoption of threat intelligence to refine those assessments. We might well find that IoT becomes a driver for the market to develop threat-intelligence platforms that can be semi-automatically integrated with IoT systems, in order that the run-time risk assessment can be enabled.

## 4.3 The challenge of understanding the glue

We focus now on understanding the harm component of risk assessment, which is so essential to our ability to prioritise and treat risk. It is clear that IoT, and the wide range of devices and actors that will form part of current and future environments, will create a vast array of connections. These are in essence the *glue* that not only enables the communications, but also IoT's many advantages: better information, greater awareness of environments, and ability to take higher quality actions quicker.

We need to be cognisant of the fact that it is not only in the protocols and communications standards that this glue is enacted, but it is also in the inner workings of the actors themselves. How they process the data they receive, and how they respond and act upon it, in itself will create effects that are inputs to other IoT actors. This is an opportunity for influence. If one can predict how these various layers of glue and behaviour are delivered and/or the way in which inputs create outputs, then one can seek to exploit the glue across these different dimensions, potentially for malicious purposes. Unfortunately, there are no existing risk assessment practices that seek to take account of this glue.

## 4.4 Failure to consider assets as an attack platform

One key failing in current risk assessment methods is that assets are only considered to be of value (and thus, to be protected), and not also from the perspective of an attack platform. There will be examples of organisations considering this, and they might well assess the risk to themselves from such asset take-overs as being driven from the possible regulatory fines (if there are any). Some organisations may also seek to quantify the intangible costs to brand, should such a situation emerge and become known to stakeholders. We raise this particularly in the context of IoT since the IoT environment brings many new devices and actors into the systems of organisations. Some of these may well be attacked and used as distributed cyber-weapons, if they can be taken over. This is an extension to issues such as insider threats, which also pose several unique challenges to IoT risk assessment [7].

Table 1: Summary of the reasons why current risk assessment approaches are inadequate for IoT

| Reason | Context |
|---|---|
| Shortcomings of periodic assessment | Current risk assessment approaches are based on periodic assessment and assume that systems will not significantly change in a short period of time. These assumptions do not hold for the IoT, where there is vast variability in scale of systems, dynamism and system coupling. |
| Changing systems boundaries yet limited system knowledge | To adequately assess a system, existing risk assessments typically mandate some reasonable knowledge (on assets, threats, probabilities of attack, potential impacts etc.). Such knowledge is extremely challenging to attain within IoT systems. Moreover, limited system knowledge means that as we enumerate risks we are likely to miss some, which in turn means that high risks could be missed entirely or mistakenly qualified. |
| The challenge of understanding the glue | Traditional risk assessment is targeted towards well-known assets, including information, devices, communication platforms and interfaces. The weakness of such a directed focus within IoT is the failure to also consider and assess: the processes through which devices are bound; the connections that allow them couple and operate; and the inner workings of the actors themselves. Each of these is a potential area of new risk. |
| Failure to consider assets as an attack platform | Within current risk assessment approaches, assets are predominately regarded as things of value to the organisation. The reality now, however, especially as it relates to IoT, is that assets (e.g., IoT devices) can be the basis for attacks. The 2016 Dyn cyberattack which involved compromised IoT devices is a perfect example of this. Prudence would therefore dictate that organisations must now accommodate for these new types of risks in their assessment processes. |

## 5. The need for new approaches to assess IoT system risk

Risks to critical infrastructure, in companies or countries, are currently assessed using methodologies which were established prior to the pervasive coupling of digital, cyber-physical and social systems. Though these methodologies are already known to have their weaknesses [6], as systems complexity and automation increases, we create new opportunities for failures which have knock-on effects through these highly connected systems. The challenge with IoT and similar coupled systems is that the periodic and knowledge-extensive processes employed by existing risk assessment approaches, may be ineffective in the face of highly dynamic systems. IoT systems are simply too fast paced for such a heavy approach.

It may be that the very philosophy at the heart of the risk assessment approaches is flawed, because the key elements and relationships are, at their heart, driven from a defensive position. Current approaches often inherently assume a single system and how it might be attacked, and the potential for resulting risk relating to the assets in question. This lens seriously fails to reveal the risks in IoT systems.

Furthermore the field is changing rapidly but our risk assessment practices are not keeping pace: viewing assets as a potential attack platform is a perfect example of this.

We therefore anticipate the need for automated and continuous risk assessment approaches, as well as the development of new support tools to assist with simulation and modelling for enhancing our predictive powers. These will take inspiration from proposed automated techniques (e.g., [3]) combined with research into risk analysis in inter-dependent systems. The core aim will be accommodating all IoT dynamics. For instance, new approaches would need to consider the potential variability of relationships, and that some may become highly (or less) trusted and that could change the risk control behaviours that surround them. The relationships and the variability of trust for the range of systems in Figure 1 will need to be anticipated and considered in the context of *potential* for risk propagation and harm.

The glue that binds the IoT systems and their actors will provide a mechanism for risk propagation, and creation of harm at physical, social (especially in the context of social machines) and economic scales. As such, the IoT actor or device, if repurposed, may be capable of facilitating harms that might be far beyond the expected. This too will need to be taken account of in a new risk assessment approach, as well as the inability of periodic assessment to respond to dramatic changes in IoT environments. It is likely that a form of run-time, near real-time, risk assessment support will be required. This could engage in more predictive considerations that aim to take account of the dynamics and changes to provide early warning of emerging risk potential. Our intention is to create such a methodology through close collaboration between industry and research.

**Acknowledgments**
This research was conducted as a part of the PETRAS Internet of Things Research Hub, a consortium of nine UK universities. The Hub is funded by the EPSRC and partner contributions, and runs in collaboration with IoTUK.

**Biographies**

*Jason R.C. Nurse is Senior Research Fellow in the Department of Computer Science at the University of Oxford and a JR Fellow at Wolfson College, Oxford. He also holds the role of Visiting Fellow in Defence & Security at Cranfield University. Jason has been selected as a Rising Star in research as a part of the UK's EPSRC RISE awards campaign, for his research into cybersecurity and privacy. His research interests include the Internet-of-Things, corporate information security, risks to identity security and privacy in cyberspace, information trust, human factors of security, and services security. Contact him at jason.nurse@cs.ox.ac.uk.*

*Sadie Creese is Professor of Cybersecurity in the Department of Computer Science at the University of Oxford. She is Director of the Global Centre for Cyber Security Capacity Building at the Oxford Martin School, and a co-Director of the Institute for the Future of Computing at the Oxford Martin School. She is engaged in a broad portfolio of cyber security research spanning identity security, situational awareness, visual analytics, risk propagation and communication, threat modelling and detection, network defence, dependability and resilience, and formal analysis. Contact him at sadie.creese@cs.ox.ac.uk.*

*David De Roure is Professor of e-Research at University of Oxford and Director of the Oxford e-Research Centre. He works at the interdisciplinary intersection of digital methods for the humanities and social sciences, including Web Science and the Internet of Things, and is a member of Cyber Security Oxford. He has strategic responsibility for Digital Humanities at Oxford, and has been Strategic Advisor to the UK Economic and Social Research Council in the area of new forms of data and real time analytics. David has extensive experience in hypertext, Web, Linked Data, and Internet-of-Things. Contact him at david.deroure@oerc.ox.ac.uk.*